\documentclass[aps,prl,twocolumn,floatfix]{revtex4}
\usepackage{graphicx,xspace,units}

\usepackage{amsmath,amssymb}

\newcommand{\micron}{\ensuremath{\unit{\mu m}}\xspace}
\renewcommand{\vec}[1]{\boldsymbol{#1}}

\begin{document}

\title{Manipulation and assembly of nanowires with
  holographic optical traps}

\author{Ritesh Agarwal,$^1$ Kosta Ladavac,$^2$ Yael Roichman,$^2$
Guiha Yu,$^1$, Charles M. Lieber, $^{1,3}$ and David G. Grier$^2$}
\address{$^1$Department of Chemistry and Chemical Biology, Harvard
  University, Cambridge, MA 02138\\
  $^2$Department of Physics and Center for Soft Matter Research,
New York University, New York, NY 10003\\
  $^3$Division of Engineering and Applied Science, Harvard University,
Cambridge, MA 02138}

\begin{abstract}
We demonstrate that semiconductor nanowires measuring just a few
nanometers in diameter can be translated, rotated, cut, fused and organized
into nontrivial structures using holographic optical traps.
The holographic approach to nano-assembly allows for simultaneous
independent manipulation of multiple nanowires, including relative
translation
and relative rotation.
\end{abstract}

\maketitle


Semiconductor nanowires
\cite{morales98,hu99a} 
are emerging as versatile building blocks for
the assembly and fabrication of a wide range of nanoelectronic
and nanophotonic devices \cite{lieber03,samuelson03,wang04}.
To date, the properties of simple nanowire-based devices 
have been determined
using nanowires deposited on the surface of a substrate
either at random or else by directed assembly
controlled by flowing fluids or electric fields
\cite{huang01,duan01,whang03}.
These latter approaches represent a significant advance over random
assembly, yet remain limited in that the end-to-end registry and
three-dimensional (3D) orientation of nanowires are not controlled,
thus precluding the rational assembly of more complex architectures
with interesting and potentially useful functional properties.  
Here we describe the use of holographic optical traps (HOTs)
\cite{dufresne98} as a general approach for parallel manipulation
and assembly of nanowires in 3D.
The HOT technique can create hundreds of independently
controlled optical traps that can manipulate mesoscopic objects 
in 3D \cite{curtis02,polin05}.
We demonstrate that cadmium sulfide (CdS) 
nanowires with cross-sections at least as small as
20~\unit{nm}
can be isolated, translated, rotated and deposited onto a substrate
with HOT arrays.
We also exploit
spatially localized photothermal and photochemical processes
induced by the well-focused traps
to cut nanowires and to fuse junctions.
These capabilities have been used to assemble nontrivial structures,
thus demonstrating the substantial potential for assembling and
subsequently investigating the functional properties of complex and
previously inaccessible structures.

\begin{figure}
  \centering
  \includegraphics[width=\columnwidth]{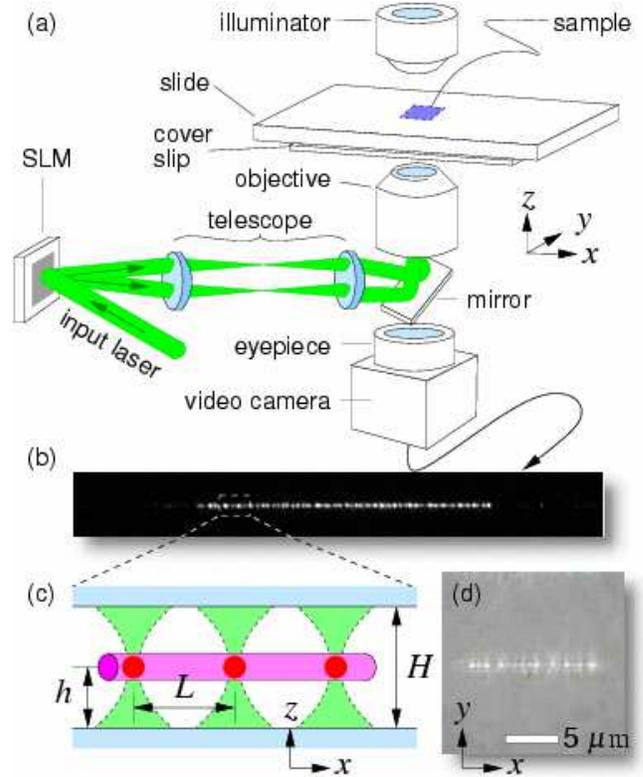}
  \caption{Holographically trapping semiconductor nanowires.
    (a) The light from a frequency-doubled solid-state laser is
    imprinted with a computer-generated hologram by a phase-shifting
    spatial light modulator (SLM) before being relayed to the
    input pupil of a high-numerical-aperture objective lens, which
    focuses the light into an array of optical traps, shown in (b).
    (c) An individual
    semiconductor nanowire can be localized by multiple optical traps, whose
    intersection with the wire typically is visualized by intense
    laser-induced fluorescence, as in (d). 
    }
  \label{fig:schematic}
\end{figure}

We synthesized CdS nanowires by the laser ablation technique
via the gold catalyzed vapor-liquid-solid (VLS) growth mechanism
\cite{morales98}.  The nanowires range from 50 to 150~\unit{nm} in
diameter with lengths ranging from 10 to 40~\micron.  The nanowires
were then suspended in ethanol by mild sonication.
Deionized water was added to the suspension (20\% v/v) prior to the
experiments to prevent the rapid evaporation of the solution, which
can lead to the deposition of nanowires on the bottom glass surface.
These samples then were charged into slit pores roughly 40~\micron
thick formed by bonding the edges of \#1 glass coverslips to the
surfaces of microscope slides.

Sealed nanowire samples were mounted for observation and manipulation
on the stage of a Nikon TE-2000U microscope
outfitted with a $100\times$ NA 1.4 Plan Apo oil-immersion objective.
This lens is used both to collect bright-field images of the dispersed nanowires
and also to focus light from a continuous wave (CW)
frequency-doubled Nd:YVO$_4$ laser operating at 532~\unit{nm}
(Coherent Verdi) into optical traps (Fig.~\ref{fig:schematic}(a)).
To create a large number of diffraction-limited optical traps, we
utilized the holographic optical tweezer (HOT) technique, as described
previously.  Our implementation uses a liquid crystal spatial light
modulator (SLM) (Hamamatus X8260 PPM) to imprint a computer-designed
phase-only hologram encoding the desired array of traps \cite{polin05}
onto the laser beam's wavefronts.
Each trap in the array can be translated independently in three
dimensions by projecting a sequence of holograms encoding the sequence
of intermediate trapping patterns.  Interactive assembly was performed
with
a BioRyx 200 holographic optical trapping system (Arryx, Inc.), also operating at
532~\unit{nm}, with an integrated MicroPoint 
pulsed laser cutter (Photonic Instruments)
operating at 440~\unit{nm}.

In our approach, nanowires dispersed in a fluid medium on the
stage of a light microscope are organized into structures
by projecting computer-designed 
patterns of diffraction-limited optical traps using
the dynamic HOT technique 
\cite{dufresne98,curtis02,polin05} (Fig.~\ref{fig:schematic}).
HOT micromanipulation relies on a generalization of the single-beam
optical gradient force traps known as optical tweezers
that can capture mesoscopic objects in 3D \cite{ashkin86}.
An individual optical tweezer is not effective for trapping
highly asymmetric structures, however,
and appears to be incapable of moving our semiconductor nanowires at
laser powers below 0.1~\unit{W}.
Increasing the power to increase the trapping force 
also induces rapid heating and 
the evolution of vapor bubbles whenever the
focal point passes through a nanowire, and to visible changes in the
nanowires themselves, including bending, formation of nodules, and
even scission.
This is consistent with heating due to absorption in the substantial
photon flux passing through the micrometer-scale focal volume.

\begin{figure}
  \centering
  \includegraphics[width=\columnwidth]{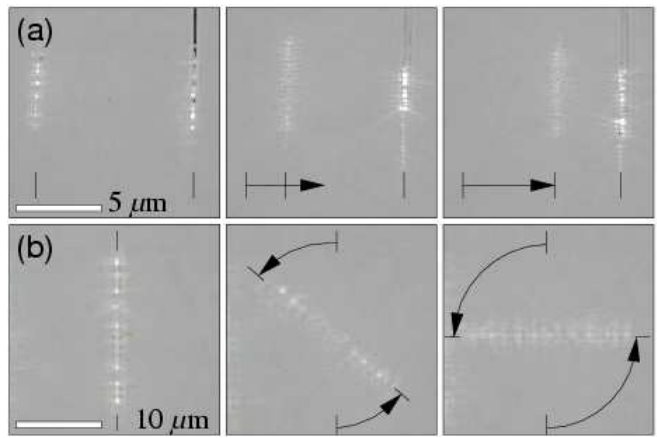}
  \caption{Translation and rotation of semiconductor nanowires by 
    holographic trap arrays.  (a) 
    Two free-floating semiconductor nanowires translated 
    toward each other with parallel arrays of holographic optical
    traps.  One wire is held stationary in one line of traps while the
    other is translated by moving a second line of traps in discrete
    steps of 700~\unit{nm}.  The traps in each line are separated by
    0.4~\micron and each trap is powered by 3~\unit{mW}.
    (b) Rotating a semiconductor nanowire by rotating an array of
    traps in discrete steps of 5$^\circ$.  The optically trapped
    CdS nanowires in these sequences appear bright because of
    photoluminescence excited by the strongly focused optical traps.
    Because these images are created with a filter that blocks the bandgap
    emission of CdS \cite{collins59}, the luminescence can be
    attributed to emission from defect sites in the CdS material \cite{bagnall00}.
  }
  \label{fig:transrot}
\end{figure}

To exert more force on the nanowires while minimizing radiative damage, we
project large numbers of holographic optical traps along the length
of each nanowire.
The image in Fig.~\ref{fig:schematic}(d)
shows a freely floating
semiconductor nanowire ca.~15~\micron long 
captured by an array of
holographic optical traps with an inter-trap
separation of 0.4~\micron.
Once aligned and localized in the array of traps, the nanowire 
can be translated at speeds up to $u = 10~\unit{\micron/sec}$ 
by moving the array across
the field of view (e.g., Fig.~\ref{fig:transrot}(a))
or by moving the sample stage relative to the array.
This upper bound can be used to estimate the optical trapping force.
The drag on a cylinder of length $L$ and radius $a$
traveling through an unbounded fluid of viscosity $\eta$ at low
Reynolds number is
\cite{batchelor70a}
\begin{equation}
  \label{eq:F0}
  F_\infty = 4 \pi \eta \, 
  \left(\epsilon + 0.193 \epsilon^2 + 0.215\epsilon^3\right) \, L \, u,
\end{equation}
where $\epsilon = [ \ln(L/a) ]^{-1}$.
This sets a lower limit on the optically applied force of
$0.2~\unit{fN/trap}$ for the ca.~80~\unit{nm} diameter CdS
nanowire used in this measurement.
The actual drag is substantially enhanced by the
need to satisfy no-flow boundary conditions at the 
nearby coverslip, which
is $h \approx 0.5~\micron$ away from the nanowire's center.
To lowest order in $a/h$, the corrected drag is \cite{takaisi55},
\begin{equation}
  \label{eq:F}
  F(h) = \frac{F_\infty}{\ln \left(\frac{2h}{a}\right)},
\end{equation}
which would increase the estimate for the trapping force by at least a
factor of two.

Although these estimates suggest that a single optical tweezer should
be able to manipulate a nanowire, a point-like trap's symmetry allows
a nanowire to rotate into an orientation that minimizes drag, and thus
to escape from the trap.
The spatially extended trapping potential 
provided by the holographic optical tweezer
array maintains the nanowire's orientation and thus makes 
controlled translation possible.
As few as two traps can capture and translate a nanowire, although
more stable trapping is observed for multiple traps arranged in a line.
Comparable trapping and orientation control has been demonstrated
for single CuO nanorods \cite{yu04} in a linear optical tweezer created
with a cylindrical lens.
Our HOT approach offers the additional benefit of manipulating
multiple nanowires simultaneously and independently in complex
ways, as described below.

Figure~\ref{fig:transrot}(a) and the associated video show two
CdS nanowires being manipulated by two arrays of traps projected
simultaneously with a single computer-generated hologram.
One nanowire is held stationary while the second is advanced
in steps of 0.7~\micron by projecting an appropriately
designed sequence of holograms at 1~\unit{s} intervals.
Similar sequences also can be used to rotate
a nanowire precisely, as shown in Fig.~\ref{fig:transrot}(b).
The video of this process demonstrates 
that both the separation and relative
orientation of two
nanowires can be controlled in this way, thereby providing the
two basic capabilities required for building complex architectures.

\begin{figure}
  \centering
  \includegraphics[width=\columnwidth]{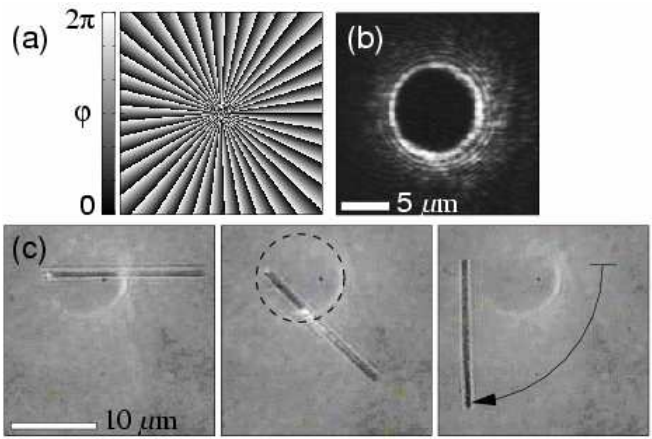}
  \caption{Rotating a semiconductor nanowire with the orbital angular
    momentum 
    flux of a helical mode of light.  
    (a) When transmitted to the SLM,
    the helical phase mask
    $\varphi(r,\theta) = \ell \theta$ transforms the wavefronts of a
    TEM$_{00}$ laser mode into an $\ell$-fold
    helix.  This helical beam focuses into the ring-like optical trap,
    shown in (b).
    The orbital angular momentum density in this trap 
    can be used to rotate a semiconductor nanowire, as shown in
    the sequence of photographs in (c), which are separated by
    1~\unit{sec} intervals.  The dashed circle in shows the position
    of an $\ell = 30$ optical vortex at 1~\unit{W}.}
  \label{fig:vrotate}
\end{figure}

The phase holograms used to create holographic optical traps also
can modify the individual beams' wavefronts to create
optical micromanipulators that do not require active updating
to process nanowires.
Specifically, a single static optical tweezer can be
transformed into an
optical vortex \cite{he95,simpson96,gahagan96,curtis03} by imposing
a helical phase profile $\varphi(\vec{r}) = \ell \theta$ onto the
trapping laser's wavefront.  
Here, $\vec{r} = (r,\theta)$ 
is a polar coordinate transverse to the
beam's axis and $\ell$ is an integer winding number defining the
wavefronts' helicity.
The effect of this modulation is to transform a point-like optical
tweezer into a ring-like trap whose radius scales linearly with
winding number \cite{curtis03,sundbeck05}, and whose photons each
carry an orbital angular momentum, $\ell \hbar$, in addition to
their intrinsic spin angular momentum \cite{allen92},
that can be transferred to objects illuminated by the
ring of light \cite{he95a,simpson97,oneil02,curtis03}.
The resulting torque causes the nanowire in Fig.~\ref{fig:vrotate}
to rotate,
even though the trap itself is static.
Arrays of optical vortices can be used to rotate large numbers
of nanowires rapidly in parallel, although with less precise angular
control than dynamic arrays of conventional optical tweezers.

\begin{figure}
  \centering
  \includegraphics[width=\columnwidth]{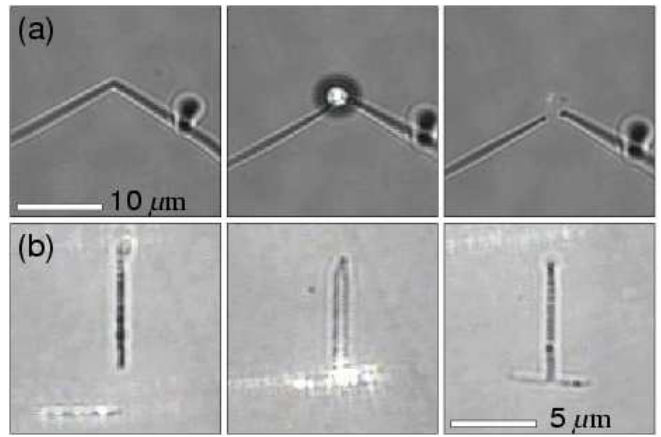}
  \caption{Transforming nanowires with intense focused beams of light.
  (a) Cutting a semiconductor nanowire with an optical
  scalpel. A bent nanowire is brought to the focus of an
  optical trap powered by 0.5~\unit{W}. An exposure time of
  100~\unit{ms} results in a clean cut at the bend.
  (b) Fusing two semiconductor nanowires into a
  free-floating assembly.
  The two nanowires are first trapped and then manipulated to form
  a T-junction.  An optical trap powered by 100~\unit{mW} is then
  focused on the junction for 1~\unit{s} to non-destructively fuse the
  wires.  The T-junction then floats freely once the traps are extinguished.}
  \label{fig:cutfuse}
\end{figure}

We also have used HOT arrays to investigate other modes of
manipulation
that could be important for assembling complex structures.
First, a trapped nanowire can be translated along the optical axis
to the surface of a substrate.
If the nanowire has not been stabilized, for example with surfactant,
this causes the nanowire to be deposited irreversibly through its
van der Waals interaction with the substrate.
In cases where the nanowires are stabilized, increasing the laser
power
in the trap array still can yield irreversible and site-specific
deposition of nanowires with controlled orientation.
Second, tightly focused optical traps at higher powers can be
used to cut nanowires, as shown in Fig.~\ref{fig:cutfuse}(a).
Here, a 0.5~\unit{W} CW optical tweezer focused on a nanowire for
ca.~100~\unit{msec} acts as an optical scalpel.  Finer cuts requiring
substantially less power can be achieved with short laser pulses 
at shorter wavelengths \cite{joglekar04}.

Once nanowires have been cut to length and organized into specific
configurations, forming junctions between them
is critical for transforming these structures into 
electronic and photonic devices of the types that have been
recently investigated and proposed
\cite{lieber03,duan01}.
The HOT approach opens up new opportunities for creating such
junctions.
For example, the translation and rotation operations can be used
to assemble two freely diffusing nanowires into a T-junction.
Applying a high-power pulse (100~\unit{mW}, 1~\unit{s}) irreversibly
fuses the nanowires to form a rigid T-junction that freely diffuses
in solution when the HOTs are removed (Fig.~\ref{fig:cutfuse}(b)).  
These results highlight further the
power of our approach; that is, it can be used to translate and rotate
nanowires in a reversible manner, and also to irreversibly modify
them through site-specific fusion, deposition and cutting.

\begin{figure}
  \centering
  \includegraphics[width=\columnwidth]{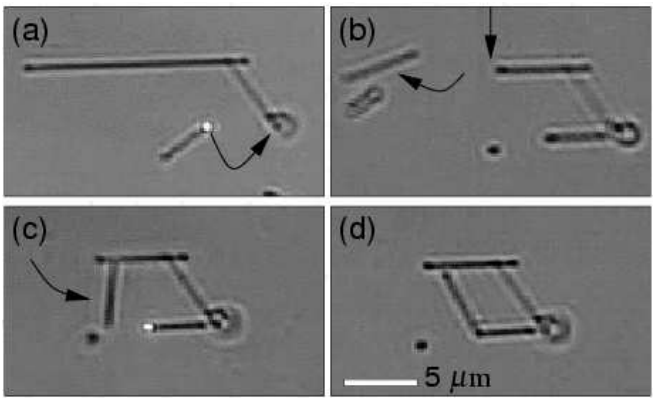}
  \caption{Assembly of rhombus constructed from semiconductor nanowires using
  holographic optical traps.  (a) A nanowire is translated towards an
  existing structure created earlier by trapping and fusing two
  nanowires.
  (b) The long nanowire is then cut with a pulsed optical scalpel.  
  (c) The resulting free-floating nanowire
  piece then is brought back to the partially completed structure.  
  (d) The structure is completed by fusing both ends of the fourth nanowire.}
  \label{fig:rhombus}
\end{figure}

Lastly, we have combined all of the manipulation steps described
above to assemble a substantially more complex structure, as shown
in Fig.~\ref{fig:rhombus}.
This interactive assembly was performed with a BioRyx 200 holographic
optical trapping system.
Figure~\ref{fig:rhombus}(a) shows a nanowire segment being translated
toward a pair of fused nanowires held in an optical tweezer
array.  
After being translated and rotated into
position, the additional segment is fused to the larger structure with a
0.5~\unit{W} pulse of light distributed over 10 traps lasting 2~\unit{s}.
Next, the longer
nanowire
in the partially completed structure is cleanly cut
(Fig.~\ref{fig:rhombus}(b)) with a short-wavelength laser pulse
($100~\unit{\mu J}$, $\lambda = 440~\unit{nm}$, $5~\unit{ns}$).
The resulting free-floating nanowire segment is captured with multiple
traps, and brought back to the optically trapped structure
(Fig.~\ref{fig:rhombus}(c)) to form a rhombus.
Finally, additional laser pulses fuse the nanowires into a stable
closed structure (Fig.~\ref{fig:rhombus}(d)).

In summary, the results presented here demonstrate that holographically projected
arrays of optical traps can
be used to manipulate and assemble semiconductor nanowires into precisely organized
two-dimensional and three-dimensional structures.
In the future, it should be possible to optimize this process
by tuning the laser wavelength to enhance the optical
trapping force.
The approach also will become substantially faster and more highly
parallel
with advances in holographic trapping technology.  
Optical assembly of functional subunits will
facilitate hierarchical fabrication of larger systems, 
through processes that might exploit
complementary techniques such as chemically-directed self-organization.
The HOT technique also can be extended to bring together diverse
nanoscale building blocks such as nanotubes \cite{plewa04} or
nanoparticles \cite{ajito02},
to utilize their unique properties in conjunction with those of
nanowires.
In addition, dynamic systems can be created by exploiting the
dynamically configurable nature of optical traps.
We believe that the exciting opportunities provided by the HOT
technique
for nanofabrication with unprecedented and exquisite spatial control
will be crucial for creating integrated and functional nanosystems in
the future.

This work was supported by the National Science Foundation
  (DBI-0233971 and DMR-0450878) and Defense Advanced Research Projects
  Agency (N00014-04-1-0591; GA9550-05-1-0444).


\end{document}